\documentclass[twocolumn,pre,showpacs,superscriptaddress,floatfix]{revtex4-1}
\usepackage{graphicx}
\usepackage{amsmath}
\usepackage{amssymb}
\usepackage{color}
\usepackage{bm}

\begin{document}
\title{Nonequilibrium steady states in Langevin thermal systems}

\author{Hyun Keun Lee}
\affiliation{Department of Physics, Sungkyunkwan University, Suwon 16419, Korea}
\author{Sourabh Lahiri}
\affiliation{School of Physics, Korea Institute for Advanced Study,
Seoul 02455, Korea}
\author{Hyunggyu Park}\email{hgpark@kias.re.kr}
\affiliation{School of Physics, Korea Institute for Advanced Study,
Seoul 02455, Korea}

\date{\today}
\begin{abstract}
Equilibrium is characterized by its fundamental properties such as the detailed balance, the fluctuation-dissipation relation, and no heat dissipation. Based on the stochastic thermodynamics, we show that these three properties are equivalent to each other in conventional Langevin thermal systems with microscopic reversibility. Thus, a conventional steady state has either all three properties (equilibrium) or none of them (nonequilibrium). In contrast, with velocity-dependent forces breaking the microscopic reversibility, we prove that the detailed balance and the fluctuation-dissipation relation  mutually exclude each other and no equivalence relation is possible between any two of the three properties. This implies that a steady state of Langevin systems with velocity-dependent forces may maintain some equilibrium properties but not all of them.
Our results are illustrated with a few example systems.
\end{abstract}


\maketitle

\section{Introduction}

The detailed balance (DB)~\cite{gardiner,kampen} and the fluctuation-dissipation relation (FDR)~\cite{ER} are important notions in statistical mechanics. In equilibrium, both are known to be valid and also play crucial roles in understanding near-equilibrium systems perturbed by a small external force. The FDR connects a linear response to the associated equilibrium correlation~\cite{lrt}, while the DB is responsible for the Onsager reciprocal symmetry~\cite{orr}. The linear response theory equipped with these properties~\cite{lrt} is a powerful tool in probing various physical properties like thermal/electric conductivities, compressibility, and so forth near equilibrium.

Beside the DB/FDR, no heat dissipation (NHD) is also known as an equilibrium property~\cite{callen}. As the entropy production is given by the Clausius form, the NHD implies total entropy conservation. These three equilibrium properties (DB/FDR/NHD) are expected to hold simultaneously in equilibrium. Then, a natural question arises whether any nonequilibrium steady state may exist with some of these properties. An usual expectation is that none of these properties would hold in nonequilibrium steady
states~\cite{gFDR}. However, we demonstrate in the following that this is not necessarily the case in general.

\begin{figure}
\includegraphics*[width=\columnwidth]{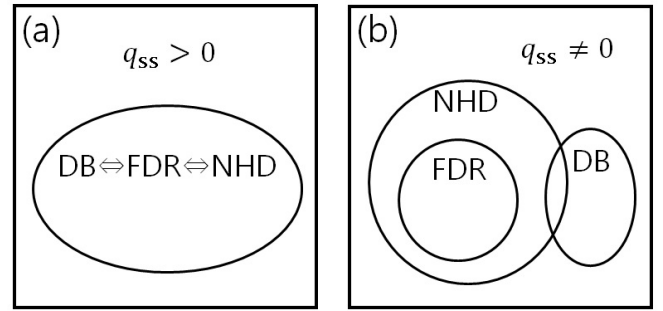}
\caption{Classification of steady states in the absence/presence of a velocity-dependent force (a)/(b). Each domain is labelled by its property  and $q_{\rm ss}$ represents the heat-dissipation rate in a steady state.}
\label{diag}
\end{figure}

In this paper, we first show the equivalence of the three properties (DB/FDR/NHD) in conventional Langevin thermal systems without velocity-dependent forces. This equivalence leads to the conclusion that any of the equilibrium properties  should not hold in nonequilibrium.  However, in the presence of velocity-dependent forces, we prove that the equivalence relation is violated and some nonequilibrium steady states may possess one or two equilibrium properties but not all three of them. In particular, the DB and the FDR exclude each other. This remarkable difference originates from microscopic irreversibility encoded in velocity-dependent forces. Our results are presented in a schematic classification diagram of steady states in Fig.~\ref{diag}.

The most well-known and fundamental velocity-dependent force is the Lorentz force acting on a moving charged particle by a magnetic field. The Lorentz force is, however, very special because it generates no work, so cannot change the system energy. This leads to the NHD in the steady state. Moreover, one finds the Boltzmann distribution in the steady state, which guarantees the FDR~\cite{orr,marconi2008}. This is probably why its steady state is often called {\em equilibrium}. However, due to the inherent microscopic irreversibility, the {\em standard} DB does not hold as well as the Onsager reciprocal
symmetry~\cite{lorentz} (more discussions will follow in Sec.~IV C on this point).

Recently, there has been a lot of research activities in collective motion of active/passive Brownian particles with velocity-dependent interactions~\cite{vicsek,tailleur,sevilla,noh,ams,ams2,GC}, molecular refrigerators with velocity-dependent forces through a feedback process (cold damping)~\cite{Kim0407,sano2011,jun,rosinberg,Hao2014,Kwon2016,Um,jourdan,cohadon,sagawa}, and nonlinear velocity-dependent friction (Sisyphus cooling)~\cite{barkai}. Furthermore, there has been efforts to understand bio matter dynamics and measure its {\em nonequilibriumness} by examining the violation of the FDR~\cite{mizuno2007,cugliandolo,tailleur2015,wijland2015,cates}. Therefore, it is imperative and also timely to re-examine all three equilibrium properties in systems with velocity-dependent forces  in a general framework.

In this study, we focus on underdamped Langevin systems in contact with a single thermal reservoir, but the generalization and application to overdamped systems as well as some bio matter systems~\cite{sevilla,noh,cates} are also possible. However, it is still nontrivial to generalize our discussions to systems with {\em athermal} noises associated with a velocity-dependent force in many active matter systems, which is left for future study.

\section{Langevin thermal systems}

We consider an underdamped Brownian motion of a particle of mass $m$  in contact with a heat reservoir.
Let ${\vec x}(t)=\{x_\mu(t)\}$ be the position vector of the particle at time $t$. The Langevin equation describing
its evolution reads
\begin{equation}
\label{ole0}
m\dot {\vec v}(t) = {\vec g}({\vec x}(t),{\vec v}(t))-\gamma {\vec v}(t)+\vec{\xi}(t)~,
\end{equation}
where ${\vec v}(t)\equiv \dot {\vec x}(t)$ is the velocity, ${\vec g}({\vec x}(t),{\vec v}(t))$ stands for the sum of the conservative and non-conservative mechanical forces, and $-\gamma {\vec v}(t)+{\vec \xi}(t)[\equiv {\vec f}_{\rm res}(t)]$ is the reservoir force. ${\vec \xi}(t)=\{\xi_\mu(t)\}$ is the gaussian white noise, satisfying
$\langle \xi_\mu(t)\rangle=0$ and $\langle {\xi_\mu(t)\xi_\nu(t')} \rangle = 2\gamma T \delta_{\mu\nu}\delta(t-t')$
with the reservoir temperature $T$ (we set the Boltzmann constant $k_B=1$). This Langevin equation can also describe the evolution of interacting many particle systems, if one consider ${\vec x}(t)$ as a collection of position vectors.

For simplicity, we present our results for a single particle system in one spatial dimension.
Generalization to interacting many particle systems in higher spatial dimensions is straightforward.
Thus, the Langevin equation we will study is
\begin{equation}
\label{ole}
m\dot v(t) = g(x(t),v(t))-\gamma v(t)+\xi(t)~.
\end{equation}
The mechanical force $g$ is divided into the
even and odd parts; $g=g^{\rm e}+g^{\rm o}$, such that $g^{\rm e}(x,v)=g^{\rm e}(x,-v)$ and
$g^{\rm o}(x,v)=-g^{\rm o}(x,-v)$. In the conventional case without a velocity-dependent force ($g=g(x)$),
no odd part exists ($g^{\rm o}=0$). With velocity-dependent forces, the odd part must exist
at least as the highest order term in $v$, in order to secure the stability of a steady state. For example, when $g\propto v^2$ without $-v^3$ term,
one can easily show that $\langle v(t)\rangle$ diverges in the long-time limit and a steady state does not exist.
This can be checked explicitly through the Fokker-Plank formulation~\cite{gardiner,kampen}.

It is important to notice that the odd part $g^{\rm o}$ breaks the microscopic reversibility (the symmetry of the deterministic equation of motion under time reversal). This microscopic {\em irreversibility} is the key feature, originated from
 the presence of velocity-dependent forces, which makes the steady-state diagram completely different from the
 conventional case in Fig.~\ref{diag}.

\section{Steady-state properties }

In this section, we investigate how the steady-state properties
such as the detailed balance, entropy production, heat dissipation, and fluctuation-dissipation relation
are interrelated in the general Langevin thermal systems.

\subsection{detailed balance}

The detailed balance (DB) condition is the probabilistic current balance between all pairs of microscopic states $\sigma=(x,v)$
in the steady state.
This can be written, in terms of the transition rate $\Gamma(\sigma'|\sigma)$ from $\sigma$ to $\sigma'$ and the steady-state distribution $p_{\rm ss}(\sigma)$, as~\cite{gardiner}
\begin{equation}
\label{db}
\Gamma(\sigma'|\sigma)p_{\rm ss}(\sigma) = \Gamma(\epsilon\sigma|\epsilon\sigma')p_{\rm ss}(\epsilon\sigma')~,
\end{equation}
where the parity operator $\epsilon$ acts as $\epsilon\sigma=(x,-v)$. Thus, the DB condition ensures the stochastic time-reversal symmetry in the steady state.
Furthermore, one can derive the steady-state parity symmetry as $p_{\rm ss}(\epsilon\sigma)=p_{\rm ss}(\sigma)$ from the
DB condition in Langevin systems~\cite{Kwon2016,Yeo,gardiner}. Note that, for systems described by a discrete jumping Markov process, this parity symmetry is an additional independent condition for equilibrium~\cite{Lee2013}.

With the DB condition in the steady state, one can derive a relation between two-time correlators of arbitrary observables, ${\cal O}_1$ and ${\cal O}_2$, in the steady state as
\begin{equation}
\label{corr}
\langle {\cal O}_1(\sigma(t')) {\cal O}_2(\sigma(t)) \rangle
=\langle {\cal O}_1(\epsilon \sigma(t)) {\cal O}_2(\epsilon \sigma(t')) \rangle~.
\end{equation}
For example, we find $\langle x(t')v(t)\rangle=-\langle x(t)v(t')\rangle$ and $\langle x(t')\dot{v}(t)\rangle=\langle x(t)\dot{v}(t')\rangle$.
The time-reversal symmetry ensured by the DB condition and the time translational symmetry in the
steady state are the key elements in deriving the above equation. Details of the derivation are found in Appendix~\ref{correlator}. We emphasize that this consequence of the DB condition is valid even in the presence of
velocity-dependent forces.

\subsection{entropy production and heat dissipation}

 Consider the total entropy production from one microscopic state $\sigma$ to another
$\sigma^\prime$ during an infinitesimal time interval $dt$ in a steady state. This can be written as
an irreversibility measure under time reversal~\cite{Sekimoto,seifert2005,Kwon2016,Yeo} such that
\begin{equation}
\label{tep}
dS^{\rm tot}_{\rm ss}=\ln \frac{\Gamma(\sigma'|\sigma)p_{\rm ss}(\sigma)}
{\Gamma(\epsilon\sigma|\epsilon\sigma')p_{\rm ss}(\sigma')}~.
\end{equation}
This is the core setup of the stochastic thermodynamics~\cite{Sekimoto,seifert2005}.
Note that the DB condition implies $dS^{\rm tot}_{\rm ss}=0$ for any transition of $\sigma \rightarrow \sigma'$
in Langevin systems.

It is also well known that
the integral fluctuation relation for this quantity always holds as~\cite{evans,Kurchan1998,seifert2005}
\begin{equation}
\label{ift}
\langle e^{-dS^{\rm tot}_{\rm ss}} \rangle =1~,
\end{equation}
which leads to the thermodynamic second law: $\langle dS^{\rm tot}_{\rm ss}\rangle \ge 0$.

In the conventional case without a velocity-dependent force, the Schnakenburg formula~\cite{snb} leads to
\begin{equation}
\label{snb}
\langle dS^{\rm tot}_{\rm ss}\rangle =
\left\langle \ln \frac{\Gamma(\sigma'|\sigma)}
{\Gamma(\epsilon\sigma|\epsilon\sigma')}\right\rangle =
\langle dQ_{\rm ss} \rangle /T=q_{\rm ss} dt/T~,
\end{equation}
where $Q_{\rm ss}$ ($q_{\rm ss}$) is heat dissipation (rate) into the reservoir.
Thus, the thermodynamic second law guarantees non-negative heat dissipation and zero entropy production is identical to no heat dissipation (NHD).
However, in the presence of a velocity-dependent force, the Schnakenburg formula does not hold
and an additional {\em unconventional} term appears as $\langle dS^{\rm tot}_{\rm ss}\rangle = q_{\rm ss} dt/T+\langle dS^{\rm unc}_{\rm ss}\rangle$~\cite{GC,Hao2014,Kwon2016}. As $\langle dS^{\rm unc}_{\rm ss}\rangle$ can be arbitrary, depending on
the form of velocity-dependent forces, the thermodynamic second law does not guarantee the non-negativeness of heat
dissipation in the steady state. In fact, $q_{\rm ss}$ is negative in the cold damping problem~\cite{Kwon2016}, which will
be discussed later.

Now, consider consequences of the DB condition in the steady state in terms of heat
dissipation. As seen in Eq.~\eqref{tep}, the DB implies $dS^{\rm tot}_{\rm ss}=0$ for any $(\sigma,\sigma')$.
Therefore, there should be NHD ($q_{\rm ss}=0$) in the conventional case, but no constraint on $q_{\rm ss}$
due to the existence of $\langle dS^{\rm unc}_{\rm ss}\rangle$ in cases with a velocity-dependent force.

It is also trivial to prove that the NHD guarantees the DB condition satisfied in the conventional case, utilizing
the integral fluctuation theorem in Eq.~\eqref{ift}. It comes from a simple mathematical fact that,
if any random variable $y$ satisfies $\langle y\rangle =0$ and $\langle e^{-y}\rangle=1$ simultaneously, then its distribution is given by a delta function as $p(y)=\delta(y)$. Therefore, the NHD condition of $\langle dS^{\rm tot}_{\rm ss}\rangle =0$ with the integral fluctuation theorem guarantees the DB by Eq.~\eqref{tep}. In the presence of a velocity-dependent force, the
NHD has no direct consequence on the DB and vice versa.

\subsection{fluctuation dissipation relation}

First, we establish a relation between fluctuation-dissipation-relation (FDR) violation and heat dissipation to the reservoir. In fact, such a relation was already reported in~\cite{Harada2005,Baiesi2014} in terms of {\it velocity}-based statistics for the conventional systems. Here, we generalize this relation to cases with a velocity-dependent force, in terms of {\em position}-based statistics which are more easily accessible in experiments~\cite{mizuno2007,Toyabe2007}.

We begin with an identity known as the Novikov's relation~\cite{marconi2008,NV65}:
\begin{equation}
\label{kidp}
\left\langle x(t') \xi(t)\right\rangle = 2\gamma T {R}(t',t)~,
\end{equation}
where $t'$ is the time for measuring the response to a weak probing force applied at $t$ and ${R}(t',t)$ is the linear position-response function. In Appendix~\ref{Novikov}, the Novikov's relation is explicitly derived for the Langevin system by the Onsager-Machlup path integral method~\cite{OM}.
Note that the causality is already encoded in Eq.~(\ref{kidp}) as a noise realization is independent of preceding events (non-anticipating property~\cite{gardiner}). Below, we set $t'>t$
without loss of generality, so $R(t,t')=0$.

One may rewrite Eq.~(\ref{kidp}) by replacing $\xi(t)$ with $f_{\rm res}(t)+\gamma v(t)$ as
\begin{equation}
\label{hk4}
\partial_t{C}(t',t) - 2 T {R}(t',t) = H(t',t)~,
\end{equation}
where the position-position correlation function is
\begin{equation}
\label{hkC}
{C}(t',t)\equiv \langle x(t')x(t)\rangle~,
\end{equation}
and the remainder is
\begin{equation}
\label{hkH}
H(t',t)\equiv -\gamma^{-1}\langle x(t')f_{\rm res}(t))\rangle~.
\end{equation}
Interchanging $t$ and $t'$ in Eq.~(\ref{hk4}) and using the causality, one also finds
\begin{equation}
\label{hk5}
\partial_{t'}{C}(t,t') =  H(t,t')~.
\end{equation}

As the standard expression of the FDR is given by $\partial_t C(t',t)=TR(t',t)$ for any $t'$ and $t$~\cite{ER,gardiner,kampen,gFDR}, a simplest measure for the FDR violation can be defined as
\begin{equation}
\label{viol}
V(t',t) \equiv \partial_t C(t',t)-TR(t',t)~,
\end{equation}
which  represents nonequilibriumness in Langevin thermal  systems.
Considering a time derivative of Eq.~\eqref{viol} in use of Eqs.~\eqref{hk4} and \eqref{hk5} with $C(t,t')=C(t',t)$, one obtains
\begin{equation}
\label{dvss}
\partial_{t'} V(t',t) = \frac{1}{2} \left[\partial_{t'} H(t',t)+\partial_{t} H(t,t')\right]~.
\end{equation}
Taking the $t'\rightarrow t^+$ limit on Eq.~(\ref{dvss}) with Eq.~(\ref{hkH}), one arrives at
\begin{equation}
\label{Vdxg}
\gamma \lim_{t'\rightarrow t^+}\partial_{t'} V(t',t)
={{\langle dQ(t)\rangle}\over{dt}}\equiv q(t)~,
\end{equation}
where the infinitesimal heat flow into the reservoir is given as $dQ(t)=-dx(t)\circ f_{\rm res}(t)$~\cite{Harada2005,Sekimoto} with the Stratonovich multiplication denoted by $\circ$~\cite{gardiner} and $q(t)$ thus stands for the average heat-dissipation rate at $t$. In $d$ spatial dimensions, $V_{\mu\nu}(t',t)$ are considered for $\mu, \nu=1,..,d$ with $R_{\mu\nu}(t',t)=\langle x_\mu (t') \xi_\nu (t)\rangle/(2\gamma T)$ and $C_{\mu\nu}(t',t)=\langle x_\mu (t') x_\nu (t)\rangle$. Then, it follows that the heat-dissipation rate $q(t)=\gamma \sum_\mu \lim_{t'\rightarrow t^+}\partial_{t'} V_{\mu\mu}(t',t)$.

In the steady state, Eq.~(\ref{Vdxg}) is identical to the Harada-Sasa relation: {$\langle v(t^+)v(t) \rangle - T \langle v(t^+)\xi(t) \rangle = \gamma^{-1} q_{\rm ss}$} with $q_{\rm ss}= \lim_{t\rightarrow \infty} q(t)$~\cite{Harada2005}. Since a stationarity is not assumed in our derivation, Eq.~(\ref{Vdxg}) is also valid in transient situations, which was also reported in~\cite{Baiesi2014}. As our derivation is based on the position statistics, the result is directly applicable to the overdamped limit described by $\gamma \dot x(t)=g(x(t))+\xi(t)$. The most important  consequence of Eq.~(\ref{Vdxg}) is that the heat dissipation rate as a FDR violation has the same expression, regardless of the existence of a velocity-dependent force. Hence, the FDR property ($V(t^\prime,t)=0$) always guarantees the NHD ($q(t)=0$), but the reverse is not necessarily true. We will show later that the reverse is true only in the conventional case without a velocity-dependent force.

Now, we examine the consequence of the DB condition in the steady state in terms of the FDR violation term $V(t',t)$.
Using Eqs.~\eqref{hk4} and \eqref{hk5} with a steady-state property $\partial_t C(t',t)+\partial_{t'} C(t,t')=0$, Eq.~\eqref{viol} can be rewritten as
\begin{equation}
\label{hkvh0}
V(t', t)= \frac{1}{2} [H(t',t)-H(t, t')] ~.
\end{equation}
Replacing $H(t',t)$ by Eq.~\eqref{hkH} with $f_{\rm res}= m\dot v-g(x,v)$, we obtain
\begin{equation}
\label{hkvh}
V(t', t)= \frac{1}{2\gamma}\left[\langle x(t')g(\sigma(t))-\langle x(t)g(\sigma(t'))\rangle\right]~.
\end{equation}
When the mechanical force depends only on position ($g(\sigma)=g(x)$), Eq.~\eqref{corr} (DB) guarantees $V(t',t)=0$ for any $t'$ and $t$, thus the FDR must hold. 

But, in general cases with velocity dependent forces, the odd-parity part
must exist ($g^{\rm o}(\sigma)\neq 0$) for the stability as discussed in Sec.~II.
Equation~\eqref{hkvh}
%
can be simplified as
\begin{equation}
\label{viola}
V(t',t)={\gamma}^{-1} \langle x(t')
g^{\rm o}(\sigma(t)) \rangle~.
\end{equation}
Hence, the key condition for the FDR is the non-existence of $g^{\rm o}(\sigma)$ such as $v$, $xv$, $v^3$, $\cdots$, representing the microscopic irreversibility in the mechanical forces.
Therefore, the FDR should be violated if the DB is satisfied in non-conventional systems with a velocity-dependent force. Its transposition is also true, so we conclude that the DB and the FDR {\em exclude} each other in cases with a velocity-dependent force. In contrast, in the conventional systems, the DB guarantees the FDR.

\subsection{interrelations between steady-state properties}

Summarizing our results, we arrive at the following conclusion. First, in the conventional case, we showed that
DB $\Rightarrow$ FDR, FDR $\Rightarrow$ NHD, NHD $\Rightarrow$ DB. This logical circle proves that all three equilibrium properties are equivalent. In  other words, if one finds any one of these properties satisfied, all other properties should follow automatically. Furthermore, it also implies that there is no nonequilibrium steady state which satisfies any of these equilibrium properties. This equivalence is rather expected, but has never been shown explicitly, up to our present knowledge. In fact, there has been a debate on whether the FDR may survive in nonequilibrium situations by examining experimental data, which was mostly cleared up only recently in~\cite{mizuno2007}.

Second, in non-conventional cases with a velocity-dependent force, we showed that DB $\Rightarrow$ no FDR, FDR $\Rightarrow$ NHD, and no relation between the NHD and the DB. Meanwhile, we find an example showing the NHD, but the FDR is violated
(anisotropic damping in Sec.~IV B). We thus conclude that any two properties are not equivalent and there can be various kinds of nonequilibrium steady states with one or two (but not all three) equilibrium properties as well as none of them. Moreover, the exclusivity of the FDR and the DB puts a strong constraint on experimental/simulation data for a response function in thermal systems with a velocity-dependent force, because the Onsager reciprocal symmetry (enforced by the DB) and the standard FDR  with the corresponding correlation function cannot be satisfied simultaneously. Note that both can be measured experimentally. Our results are summarized into schematic classification diagrams of steady states in Fig.~\ref{diag}.

\section{examples}

We present a few example systems with a velocity-dependent force to confirm Fig.~\ref{diag}(b).

\subsection{cold damping}
The cold-damping problem has been actively studied in recent years, both theoretically and experimentally, in various perspectives~\cite{Kim0407,sano2011,jun,rosinberg,Hao2014,Kwon2016,Um,jourdan,cohadon,sagawa}.
This may be the simplest example with a velocity-dependent force in contact with a single heat reservoir.
Here, we consider the one-dimensional case with $g(x,v) = -k x -\gamma'v$ with $k>0$ and $\gamma'>-\gamma$ for stability.
The velocity-dependent force $-\gamma' v$ represents an additional external friction.

The DB is shown to hold for this system in~\cite{Hao2014,Kwon2016}. Thus, according to the exclusivity, the FDR should not hold. To check this, we calculate $V(t)\equiv V(t,0)$ for $t>0$, using Eq.~\eqref{viola}. Multiplying either side of Eq.~(\ref{ole}) by $\gamma^{-1} g^{\rm o}(x(0),v(0))=-\frac{\gamma'}{\gamma} v(0)$ and taking the steady-state average, we obtain an ordinary second-order differential equation for $V(t)$ as
\begin{equation}
\label{sde}
m\ddot {V}(t) + (\gamma +\gamma')\dot {V}(t) + k V(t)=0~,
\end{equation}
which is identical to the familiar differential equation of a damped harmonic oscillator.

The steady-state distribution of this system is $p_{\rm ss}(x,v)\propto e^{-(kx^2+mv^2)/(2T_{\rm eff})}$ for $T_{\rm eff}=T\gamma/(\gamma+\gamma')$~\cite{Hao2014,Kwon2016}. Then, the initial conditions for the differential equation are given by $V(0^+)=-\frac{\gamma'}{\gamma}\langle x(0^+) v(0)\rangle=0$ and $\dot {V}(0^+)=-\frac{\gamma'}{\gamma}\langle v(0^+) v(0)\rangle=-\frac{T}{m} \frac{\gamma'}{\gamma+\gamma'}$.  With these initial values, the solution is obtained as
\begin{equation}
\label{dF}
V(t) = -\frac{T}{m}\frac{\gamma'}{\gamma+\gamma'}
\frac{e^{-\beta t}}{2\omega}
\left[e^{\omega t}-e^{-\omega t}\right]~,
\end{equation}
where $\lambda_{\pm}=-\beta\pm \omega$ are the two roots of the characteristic equation $m\lambda^2+(\gamma+\gamma')\lambda + k=0$. As expected, the FDR does not hold ($V(t)\neq 0$)~\cite{com3}. The heat flow can be calculated from Eq.~\eqref{Vdxg}, resulting in $q_{\rm ss} = -{T\over m}{{\gamma\gamma'}\over{\gamma+\gamma'}}$. Note that $q_{\rm ss}$ is nonzero and can be negative as well. Thus, the unconventional entropy term $\langle dS^{\rm unc}_{\rm ss}\rangle$ is absolutely necessary to satisfy the thermodynamic second law $\langle dS^{\rm tot}_{\rm ss}\rangle \ge 0$.
 This system belongs to the region  in Fig.~\ref{diag}(b), where the DB holds while the NHD does not.

\subsection{anisotropic damping}

Consider a 2-dimensional version of the first example like $g_\mu(x,v) = -kx_\mu - \gamma_\mu v_\mu$ $(\mu=1,2)$ with anisotropic frictions with $\gamma_1>0$ and $\gamma_2 > -\gamma$. Complete decoupling between components ($\mu=1$ and 2) trivially yields the DB again and also gives the heat dissipation $q_{\rm ss}= -\sum_\mu\frac{T}{m}\frac{\gamma\gamma_\mu}{\gamma+\gamma_\mu}$.
 Interestingly, this  may reduce to zero (NHD) with the special choice of $\gamma_2 = -{{\gamma\gamma_1}\over{\gamma+2\gamma_1}}>-\gamma$. With this choice, $\gamma_2$ is negative, so
 the external force $-\gamma_2 v_2$  inputs more energy into the 2nd direction,
 which is compensated by dissipation in the 1st direction.
 This situation can be achieved experimentally by introducing such a feedback control process by measurements.
 This example thus belongs to the overlap region between the DB and the NHD in Fig.~\ref{diag}(b).
 As expected, the FDR does not hold with $p_{\rm ss}(x,v)\propto e^{-\sum_\mu (kx_\mu^2+mv_\mu^2)/(2T^\mu_{\rm eff})}$ for $T^\mu_{\rm eff}=T\gamma/(\gamma+\gamma_\mu)$, which is not a Boltzmann distribution with the reservoir temperature $T$.

\subsection{Lorenz-like force}
We study a system with a Lorentz-like anti-symmetric force in two dimensions, as discussed early in this paper.
The mechanical force is given by $g_\mu=-k x_\mu -\sum_\nu A_{\mu\nu} v_\nu$ with the anti-symmetric matrix $\mathbf{A}=\{A_{\mu\nu}\}=\left(
\begin{matrix}
0 & B \\ -B & 0
\end{matrix}
\right)$. A charged particle under the magnetic field is a typical example.
It is clear that $q_{\rm ss}=0$ (NHD) because the Lorentz-like  force does not generate work.

We now examine the validity of the FDR and the DB by an explicit calculation as follows. To this end, we calculate both $C_{\mu\nu}(t)\equiv C_{\mu\nu}(t,0)$ and $R_{\mu\nu}(t)\equiv R_{\mu\nu}(t,0)$ in the steady state. Similar to the procedure to get Eq.~\eqref{sde}, we multiply $x_\mu (0)$ to Eq.~\eqref{ole} and take the steady-state average, yielding the ordinary second-order {\em matrix} differential equation as
\begin{equation}
\label{sdef}
m\ddot{\mathbf{C}}(t) + (\gamma \mathbf{I} +\mathbf{A})\dot{\mathbf{C}}(t)+ k \mathbf{C}(t)=0~,
\end{equation}
where the position correlation matrix $\mathbf{C}(t)=\{C_{\mu\nu}(t)\}$ and $\mathbf{I}$ is the identity matrix. The initial conditions are given by $\mathbf{C}(0^+)=(T/k) \mathbf{I}$ and $\dot{\mathbf{C}}(0^+)=\mathbf{0}$ from the steady-state Boltzmann distribution $p_{\rm ss}\propto e^{-\sum_\mu(kx_\mu^2+mv_\mu^2)/(2T)}$~\cite{orr,Kwon2016}. After a straightforward algebra, we obtain
\begin{equation}
\label{Cem}
\mathbf{C}(t)=
\left(
\begin{matrix}
{\rm Re}\Phi(t) & {\rm Im}\Phi(t) \\ -{\rm Im}\Phi(t) & {\rm Re}\Phi(t)
\end{matrix}
\right)~,
\end{equation}
where
\begin{equation}
\label{Cf}
\Phi(t) = \frac{T}{k}
\frac{e^{-\tilde\beta t}}{2\tilde\omega}
\left[
(\tilde\omega +\tilde\beta)e^{\tilde\omega t}-
(\tilde\omega -\tilde\beta)e^{-\tilde\omega t}
\right]
\end{equation}
with $\tilde\lambda_{\pm}=-\tilde\beta\pm \tilde\omega$, the two roots of the characteristic equation $m\tilde\lambda^2+(\gamma+iB)\tilde\lambda + k=0$.

Similarly, we find that the response function $\mathbf{R}(t)=\{R_{\mu\nu}(t)\}$ satisfies the same differential equation of ${\mathbf{C}}(t)$ in Eq.~\eqref{sdef}. The initial values can be calculated from Eq.~(\ref{kidp}) as $\mathbf{R}(0^+)=\mathbf{0}$ and $\dot{\mathbf{R}}(0^+)=(1/m)\mathbf{I}$~\cite{Harada2005,Baiesi2014,hkhu}. After a similar algebra as above, we find
\begin{equation}
\label{Rem}
\mathbf{R}(t)=
\left(
\begin{matrix}
{\rm Re}\Psi(t) & {\rm Im}\Psi(t) \\ -{\rm Im}\Psi(t) & {\rm Re}\Psi(t)
\end{matrix}
\right)~,
\end{equation}
where
\begin{equation}
\label{Rf}
\Psi(t) = \frac{1}{m}
\frac{e^{-\tilde\beta t}}{2\tilde\omega}
\left[e^{\tilde\omega t}-
e^{-\tilde\omega t}\right]~.
\end{equation}
Plugging Eqs.~\eqref{Cem} and \eqref{Rem} into Eq.~\eqref{viol} and using a steady-state property $\partial_t\mathbf{C}(t',t)=-\partial_{t'}\mathbf{C}(t',t)$, we get
\begin{equation}
\label{FDRnoDB}
\mathbf{V}(t)=-\dot{\mathbf{C}}(t)-T\mathbf{R}(t)=\mathbf{0}~,
\end{equation}
which is by itself the FDR. We note that the validity of the FDR
can be shown directly from the steady-state Boltzmann distribution by utilizing the generalized 
FDR~\cite{gFDR,marconi2008}.

The DB violation can be easily noticed from the nonsymmetric property of the correlation matrix ${\mathbf{C}}(t)$ in Eq.~\eqref{Cem}, such that $\langle x_1(t)x_2(0)\rangle \neq \langle x_1(0)x_2(t)\rangle$, which clearly violates the DB property in Eq.~\eqref{corr}. In fact, the DB violation is explicitly calculated in terms of the average total entropy production rate as $\langle \dot{S}^{\rm tot}_{\rm ss} \rangle=2B^2/(\gamma m)>0$~\cite{Kwon2016}.
This example manifests that the FDR without the DB is indeed possible, which belongs to the region in
Fig.~\ref{diag}(b), where both the FDR and the NHD hold, but the DB is violated.

We remark that, when  one considers a different system with $-\mathbf{A}$ instead of $\mathbf{A}$ in calculating
the right hand sides of Eqs.~\eqref{db} and \eqref{corr}, the equalities hold for this Lorentz-like (magnetic) system. These
equalities are called as the extended DB in~\cite{kampen,orr,lorentz}. The extended Onsager
reciprocal symmetry is also defined in a similar way~\cite{lrt,orr}. These relations are useful because they provide exact relations between some physical quantities of two different systems such as the Onsager reciprocal symmetry between magnetic systems with opposite magnetic fields~\cite{lrt,orr}. However, these relations are not relevant to the time reversibility of a given process~\cite{gardiner,kampen}.


\subsection{nonlinear Lorentz-like force}

Consider a more general nonlinear force generating no work such as
$g_\mu=-k x_\mu -\sum_\nu A_{\mu\nu}({\vec v}) v_\nu$. We can find such forces in
some phenomenological models, describing flocking motion of Brownian particles~\cite{sevilla,noh}.
Due to the anti-symmetry of the matrix $\mathbf{A}$,
this force does not generate work, thus no heat dissipation in the steady state (NHD).
The steady-state distribution $p_{\rm ss}$ can be studied easily through the Fokker-Planck formulation~\cite{gardiner,kampen},
which proves that $p_{\rm ss}$ takes the Boltzmann distribution only when
 the velocity-dependent force is divergenceless; $\partial_{\vec v} \cdot[\mathbf{A}({\vec v}) {\vec v}]=0$.

In the previous magnetic system where the Lorentz force is divergenceless, the steady-state distribution is
found to be Boltzmannian and thus the FDR does hold. However, in general nonlinear cases, we expect that
the FDR should not hold with a non-Boltzmann steady-state distribution. Of course, the standard DB is also violated.
This case belongs to the region in Fig.~\ref{diag}(b), where only the NHD holds.
We also note that the extended DB as well as the extended Onsager reciprocal symmetry are not satisfied unless
the velocity-dependent force is divergenceless~\cite{chunnoh}.

\subsection{other cases}

We may consider a more general
$\mathbf{A} = \left(
\begin{matrix}
0 & a \\ -b & 0
\end{matrix}
\right)$
with constants $a\neq \pm b$ in the previous linear example. From Eq.~(27) of~\cite{Kwon2016}, we can obtain $\langle \dot{S}^{\rm tot}_{\rm ss} \rangle=(a+b)^2/(2\gamma m)>0$ (broken DB) and $q_{\rm ss}=\gamma (a-b)^2/[2m(\gamma^2+ab)]\neq 0$, leading to the FDR violation by Eq.~\eqref{Vdxg}. Thus, this example belongs to the region where any of the FDR/DB/NHD does not hold. Finally, one may add a diagonal part such as
$\mathbf{A} = \left(
\begin{matrix}
\gamma' & a \\
-b & \gamma'
\end{matrix}
\right)$.
For this setting, Eq.~(27) of~\cite{Kwon2016} is again useful in finding that the DB is still violated while $q_{\rm ss}$ may vanish with a proper choice of $\gamma'$ within the stability condition. Regarding the FDR, we have checked that $\mathbf{R}(t)$ still remains in the same form of Eq.~(\ref{Rem}), while $\mathbf{C}(t)$ changes its form different from Eq.~(\ref{Cem})~\cite{hkhu}. Thus the vanishing like Eq.~(\ref{FDRnoDB}) is impossible. Consequently, with this $\mathbf A$, only the NHD holds, similar to the case with a nonlinear Lorentz-like force. Difference is that $\mathbf A$ is not antisymmetric
and not velocity-dependent. This completes scanning of all the regions in Fig.~\ref{diag}(b).

\section{Summary}

   In this paper, we show that the three equilibrium properties (FDR/DB/NHD) are equivalent to each other in the conventional systems without a velocity-dependent force (thus with the microscopic reversibility in the mechanical forces). In this case, the onset of nonequilibrium is accompanied with the simultaneous violation of the three equilibrium properties. In contrast, with a velocity-dependent force, breaking the microscopic reversibility, various kinds of nonequilibrium steady states can be realized with a partial violation of the equilibrium properties. This implies that equilibrium/nonequilibrium-ness should not be determined by examining some (not all) of the equilibrium properties. In particular, the exclusivity between the FDR and the DB is useful in probing the presence of any velocity-dependent force experimentally in a given system. The FDR guarantees the Onsager reciprocal symmetry (DB) in the conventional systems, while the FDR invalidates the Onsager reciprocal symmetry
   in the presence of a velocity-dependent force.

\begin{acknowledgments}
This research was supported by the NRF Grant No.~2015R1D1A1A01057842 (H.K.L.)
and 2017R1D1A1B06035497 (H.P.).
S.L. is supported by the Region \^{I}le de France thanks to the ISC-PIF.
\end{acknowledgments}

\appendix
\section{Two-time correlator}
\label{correlator}
We consider a two-time correlator ${\cal C}(t',t)$ in the steady state defined as
\begin{equation}
\label{apxa-1}
{\cal C}(t',t)=\langle {\cal O}_1(\sigma(t')) {\cal O}_2(\sigma(t)) \rangle~,
\end{equation}
where ${\cal O}_1 (\sigma)$ and ${\cal O}_2 (\sigma)$ are observables as functions of a
microscopic state $\sigma=(x,v)$.
This can be rewritten by a path integral as
\begin{equation}
\label{apxa-2}
{\cal C}(t',t)=\int D[\sigma]_0^T~ {\cal P}[\sigma]_0^T~ {\cal O}_1 (\sigma(t')) {\cal O}_2(\sigma(t))~,
\end{equation}
where ${\cal P}[\sigma]_0^T$ is the path probability of the system evolving along a given path
$\{\sigma(\tau)\}$ for $0\le \tau \le T$ and the integration $\int D[\sigma]_0^T$ is over all paths.
The observables are measured during the evolution ($0<t'<t<T$).

For convenience, we discretize time such that $\tau_n=n d\tau$ for $n=0,\ldots,N$ with the infinitesimal interval
$d\tau=T/N$ for large $N$. The Markovian property allows us to write the path probability as
\begin{equation}
\label{apxa-3}
{\cal P}[\sigma]_0^T = p_{\rm ss} (\sigma_0) \prod_{n=1}^{N} \Gamma(\sigma_n|\sigma_{n-1})~,
\end{equation}
where $\sigma_n=\sigma(\tau_n)$. The DB condition, Eq.~\eqref{db}, with the steady-state parity symmetry
as $p_{\rm ss}(\epsilon\sigma)=p_{\rm ss}(\sigma)$ leads to the path probability in a time-reversed form as
\begin{equation}
\label{apxa-4}
{\cal P}[\sigma]_0^T = p_{\rm ss} (\epsilon\sigma_N) \prod_{n=1}^{N} \Gamma(\epsilon\sigma_{n-1}|\epsilon\sigma_{n})~.
\end{equation}
With this form of the path probability and by changing the integral variables as $\bar{\sigma}_n\equiv \epsilon \sigma_{N-n}$,
Eq.~\eqref{apxa-2} can be rewritten as
\begin{equation}
\label{apxa-5}
{\cal C}(t',t)=\int D[\bar{\sigma}]_0^T~ {\cal P}[\bar{\sigma}]_0^T~ {\cal O}_1 (\epsilon\bar{\sigma}(T-t'))
{\cal O}_2(\epsilon\bar{\sigma}(T-t))~,
\end{equation}
where ${\cal P}[\bar{\sigma}]_0^T=p_{\rm ss} (\bar{\sigma}_0) \prod_{n=1}^{N} \Gamma(\bar{\sigma}_n|\bar{\sigma}_{n-1})$.
As $\{\bar{\sigma}_n\}$ are just dummy variables for integration, this leads to
%
%
\begin{equation}
\label{apxa-6}
{\cal C}(t',t)=\langle {\cal O}_1(\epsilon\sigma(T-t')) {\cal O}_2(\epsilon\sigma(T-t)) \rangle~.
\end{equation}
Using the time-translational symmetry in the steady state
to the right hand side of Eq.~\eqref{apxa-6},
we finally arrive at Eq.~\eqref{corr}.

Note that our derivation does not assume any conventional equilibrium property such
as the Boltzmann distribution, thus our result can apply to the general case with velocity-dependent forces.

\section{Novikov's relation}
\label{Novikov}

We start with the Langevin equation of Eq.~\eqref{ole} with a small probing force $\alpha f_{\rm p}(t)$ ($\alpha\ll 1$) as
\begin{equation}
\label{apxb-1}
m\dot v(t) = g(x(t),v(t))-\gamma v(t)+\xi(t) + \alpha f_{\rm p}(t)~.
\end{equation}
The linear position-response function $R(t',t)$ is defined through the perturbation expansion as
\begin{equation}
\label{apxb-2}
\left\langle x(t') \right\rangle_\alpha = \left\langle x(t') \right\rangle_0 + \alpha\int_{-\infty}^{\infty} dt {R}(t',t)f_{\rm p}(t) + {\cal O}(\alpha^2)~,
\end{equation}
which can be recast into the form of
\begin{equation}
\label{apxb-3}
{R}(t',t)=\frac{\delta}{\delta f_{\rm p}(t)} \left.\frac{\partial \langle x(t') \rangle_\alpha}{\partial \alpha}\right|_{\alpha=0}~,
\end{equation}
where $\frac{\delta}{\delta f_{\rm p}}$ is the functional derivative.

The average position value $\langle x(t')\rangle_\alpha$ is calculated by the Onsager-Machlup path integral~\cite{OM} as
\begin{equation}
\label{apxb-4}
\langle x(t')\rangle_\alpha=\int D[\sigma] {\cal P}_\alpha[\sigma] x(t')~,
\end{equation}
with
\begin{equation}
\label{apsb-5}
{\cal P}_\alpha [\sigma] ={\cal N} e^{-\int dt \left[\frac{1}{4\gamma T} \left(m\dot{v} +\gamma v-g-\alpha f_{\rm p}\right)^2
-\frac{1}{2}\gamma + \frac{1}{2} \partial_v g\right]}~,
\end{equation}
with the normalization constant ${\cal N}$.

By differentiating Eq.~\eqref{apxb-4}, one can easily derive the Novikov relation
\begin{equation}
\label{apsb-6}
R(t',t)=
\int D[\sigma] \frac{{\cal P}_0[\sigma]}{2\gamma T}  x(t') \xi(t) = \frac{1}{2\gamma T} \langle x(t')\xi(t)\rangle_0~,
\end{equation}
where we replace $m\dot{v}(t) +\gamma v(t)-g(x(t),v(t))$ by $\xi(t)$ from Eq.~\eqref{apxb-1} at $\alpha=0$.
Note that this relation is valid not only in the steady state but also in a transient state.
Also note that the derivation is done for any $g(x,v)$, regardless of the existence of a velocity-dependent force.



\begin{thebibliography}{99}

\bibitem{gardiner} C. Gardiner, {\it Stochastic Methods}, 4th Edition (Springer-Verlag, Berlin, 2009).
\bibitem{kampen} N. G. Van Kampen, {\it Stochastic Processes in Physics and Chemistry} (Elsevier, Amsterdam, 2007).
\bibitem{ER} A. Einstein, Annalen der Physik {\bf 322}, 549 (1905); M. Smoluchowski, {\it ibid.} {\bf 326}, 756 (1906).
\bibitem{lrt} R. Kubo, J. Phys. Soc. Jpn {\bf 12}, 570 (1957).
\bibitem{orr} L. Onsager, Phys. Rev. {\bf 37}, 405 (1931).
\bibitem{callen} H. B. Callen, {\it Thermodynamcis and an Introduction to Thermostatistics}, 2nd Edition (John Wiley \& Sons, New York, 1985).
\bibitem{gFDR} A generalized FDR was found in general chaotic/stochastic (nonequilibrium) systems~\cite{marconi2008},
in terms of an effective temperature or correlations of a priori unknown variables associated with a steady state distribution.
In this work, we retrict our discussions to the {\em standard} FDR for Langevin systems with a single heat reservoir discussed in the textbooks~\cite{gardiner,kampen}, in order to measure nonequilibriumness of steady states by its violation.
\bibitem{marconi2008} U. M. B. Marconi, A. Puglisi, L. Rondoni, and A. Vulpiani, Phys. Rep. {\bf 461}, 111 (2008); J. Prost, J.-F. Joanny, and J. M. R. Parrondo, Phys. Rev. Lett. {\bf 103}, 090601 (2009).
\bibitem{lorentz} Instead, the {\em extended} detailed balance and the {\em extended} Onsager reciprocal symmetry hold in this case~\cite{kampen,orr}.
\bibitem{vicsek} T. Vicsek, A. Czir\'ok, E. Ben-Jacob, I. Cohen, and O. Shochet, Phys. Rev. Lett. {\bf 75}, 1226 (1995).
\bibitem{tailleur} A. P. Solon and J. Tailleur, Phys. Rev. Lett. {\bf 111}, 078101 (2013); A. P. Solon, H. Chat\'e, and J. Tailleur, Phys. Rev. Lett. {\bf 114}, 068101 (2015).
\bibitem{sevilla} V. Dossetti and F. J. Sevilla, Phys. Rev. Lett. {\bf 115}, 058301 (2015).
\bibitem{noh} P.-S. Shim, H.-M. Chun, and J. D. Noh, Phys. Rev. E {\bf 93}, 012113 (2016).
\bibitem{ams} F. Schweitzer, {\it Brownian Agents and Active Particles} (Springer, Berlin, 2003).
\bibitem{ams2} P. Romanczuk, M. B\"{a}r, W. Ebeling, B. Lindner, and L. Schimansky-Geier, Eur. Phys. J. Special Topics {\bf 202}, 1 (2012).
\bibitem{GC} C. Ganguly and D. Chaudhuri, Phys. Rev. E {\bf 88}, {032102} (2013);  D. Chaudhuri, {\em ibid.}~{\bf 90}, {022131} (2014).
\bibitem{Kim0407} K. H. Kim and H. Qian, Phys. Rev. Lett. {\bf 93}, 120602 (2004); Phys. Rev. E {\bf 75}, 022102 (2007).
\bibitem{sano2011} S. Ito and M. Sano, Phys. Rev. E {\bf 84}, 021123 (2011).
\bibitem{jun} Y. Jun and J. Bechhoefer, Phys. Rev. E {\bf 86}, 061106 (2012).
\bibitem{rosinberg} T. Munakata and M. L. Rosinberg, Phys. Rev. Lett. {\bf 112}, 180601 (2014); M. L. Rosinberg, T. Munakata, and G. Tarjus, Phys. Rev. E {\bf 91}, 042114 (2015).
\bibitem{Hao2014} H. Ge, Phys. Rev. E {\bf 89}, 022127 (2014).
\bibitem{Kwon2016} C. Kwon, J. Yeo, H. K. Lee, and H. Park, J. Korean Phys. Soc. {\bf 68}, 633 (2016).
\bibitem{Um} C. Kwon, J. Um, and H. Park, EPL {\bf 117}, 10011 (2017).
\bibitem{jourdan} G. Jourdan, G. Torricelli, J. Chevrier, and F. Comin, Nanotechnology {\bf 18}, 477502 (2007).
\bibitem{cohadon} P.F. Cohadon, A. Heidmann, and M. Pinard, Phys. Rev. Lett. {\bf 83}, 3174 (1999).
\bibitem{sagawa} S. Toyabe, T. Sagawa, M. Ueda, E. Muneyuki, and M. Sano, Nature Phys. {\bf 6}, 988 (2010).
\bibitem{barkai} A. Dechant, D. A. Kessler, and E. Barkai, Phys. Rev. Lett. {\bf 115}, 173006 (2015).
\bibitem{mizuno2007} D. Mizuno, C. Tardin, C. F. Schmidt, and F. C. MacKintosh, Science {\bf 315}, 370 (2007).
\bibitem{cugliandolo} D. Loi, S. Mossa, and L. F. Cugliandolo, Phys. Rev. E {\bf 77}, 051111 (2008); Soft Matter {\bf 7}, 3726 (2011).
\bibitem{tailleur2015} A. Solon, M. Cates, and J. Tailleur, Eur. POhys. J. Spec. Top. {\bf 224}, 1231 (2015).
\bibitem{wijland2015} \'E. Fodor, M. Guo, N. S. Gov, P. Visco, D. A. Weitz, and F. van Wijland, Europhys. Lett. {\bf 110}, 48005 (2015).
\bibitem{cates} \'E. Fodor, C. Nardini, M. E. Cates, J. Tailleur, P. Visco, F. van Wijland, Phys. Rev. Lett. {\bf 117}, 038103 (2016).
\bibitem{Yeo} J. Yeo, C. Kwon, H. K. Lee, and H. Park, J. Stat. Mech. 093205 (2016).
\bibitem{Lee2013} H. K. Lee, C. Kwon, and H. Park, Phys. Rev. Lett. {\bf 110}, 050602 (2013).
\bibitem{Sekimoto} K. Sekimoto, J. Phys. Soc. Jpn. {\bf 66}, 1234 (1997); Prog. Theor. Phys. Suppl. {\bf 130}, 17 (1998).
\bibitem{seifert2005} U. Seifert, Phys. Rev. Lett. {\bf 95}, 040602 (2005); Rep. Prog. Phys. {\bf 75}, 126001 (2012).
\bibitem{evans} D. J. Evans, E. G. D. Cohen, and G. P. Morriss, Phys. Rev. Lett. {\bf 71}, 2401 (1993).
\bibitem{Kurchan1998} J. Kurchan, J. Phys. A {\bf 31}, 3719 (1998).
\bibitem{snb} J. Schnakenberg, Rev. Mod. Phys. {\bf 48}, 571 (1976); H. Hinrichsen, C. Gogolin, and P. Janotta, J. Phys.: Conf. Ser. {\bf 297}, 012011 (2011).
\bibitem{Harada2005} T. Harada and S.-i. Sasa, Phys. Rev. Lett. {\bf 95}, 130602 (2005); Phys. Rev. E {\bf 73}, 026131 (2006).
\bibitem{Baiesi2014} E. Lippiello, M. Baiesi, and A. Sarracino, Phys. Rev. Lett. {\bf 112}, 140602 (2014).
\bibitem{Toyabe2007} S. Toyabe, H.-R. Jiang, T. Nakamura, Y. Murayama, and M. Sano, Phys. Rev. E {\bf 75}, 011122 (2007).
\bibitem{NV65} E. A. Novikov, JETP {\bf 20}, 1290 (1965).
\bibitem{OM} L. Onsager and S. Machlup Phys. Rev. {\bf 91}, 1505 (1953).
\bibitem{com3} One may rewrite Eq.~\eqref{dF} in the form of $\partial_t C(t',t)=T_{\rm eff} R(t',t)$, which is an instance of the generalized FDR~\cite{gFDR,marconi2008}. However, this relation is not relevant in measuring nonequilibriumness.
\bibitem{hkhu} H. K. Lee and H. Park (unpublished).
\bibitem{chunnoh} H.-M. Chun and J. D, Noh, arXiv:1706.01691.
\end{thebibliography}
\end{document}